\documentstyle[12pt]{article}

\def\PL{\em Phys. Lett.}
\def\PR{\em Phys. Rep.}

\def\NPB{{\em Nucl. Phys.} B}
\def\PLB{{\em Phys. Lett.}  B}
\def\PRL{\em Phys. Rev. Lett.}
\def\PRD{{\em Phys. Rev.} D}

\def\TMP{\em Theor. Math. Phys.}
\def\MPLA{{\em Mod. Phys. Lett.} A}
\def\IJMPA{{\em Int. J. Mod. Phys.} A}
\begin{document}

\newcommand{\be}{\begin{equation}}
\newcommand{\ee}{\end{equation}}
\newcommand{\ba}{\begin{eqnarray}}
\newcommand{\ea}{\end{eqnarray}}

\begin{center}
{\bf NUCLEON SPIN PUZZLE: TEN YEARS LATER...}\\

\vspace*{0.2cm}
{\bf Vladimir A. Petrov}\\

\small
{Division of Theoretical Physics, Institute for High Energy Physics,
142284 Protvino, 
Russian Federation, e-mail: petrov@mx.ihep.su} \\
\end{center}

\vspace*{0.5cm}
\small
{This is a  brief survey of the present state of
ideas concerning the parton content of the nucleon spin after their
ten-years evolution since the discovery of the EMC.}

\vspace*{0.5cm}
In 1974 an article [1] appeared in the Physical Review in which
J.~Ellis and R.~Jaffe deduced some predictions (sum rules) concerning
structure  functions of polarized lepton-nucleon deeply inelastic
scattering.

The predictions were based on some apparently natural assumptions
like the 
negligible r\^ole  of strange quarks in the formation of the nucleon
spin.
The result did not cause a noticeable agiotage. The general attention
was
focused then on another problems.

Meanwhile some other important works were published  among which
one has to mention papers~[2] (calculation of the ``polarized''
anomalous dimension whose lowest order is $0(\alpha^2_s)$ in contrast
to usual $0(\alpha_s)$ for ``unpolarized'' anomalous  dimensions.
Later this anomalous dimension was calculated to three loops~[3].)
and
~[4] (establishing a relation of the problem to the chiral anomaly
and discussion of the r\^ole of the gluon helicity). 

A crash of thunder happened only in 1988 when the results of the
European Muon Collaboration were published~[5]. The  three-standard
deviation of the Ellis-Jaffe prediction from the data produced a
great impression and caused a flow of theoretical publications as
well as new experimental measurements both at CERN and SLAC.

A ``simple minded'' interpretation of the EMC data as an evidence
that the
nucleon spin is not  formed mainly by valence quarks was rejected by
some authors as absolutely counter intuitive, highly unnatural and,
thus, appealing for a more sophisticated consideration, which could
reconcile the data and ``natural'' ideas on composition of the
nucleon spin~[6]. Nonetheless some other theorists accepted
this interpretation arguing that at rather small distances
(corresponding to $Q^2_{EMC}\simeq 10$GeV$^2$) constituent quark
picture ceases to be relevant and the total nucleon spin is
``smeared'' over a large number of ``resolved'' quarks, antiquarks
and
gluons~[7]. They also indicated that the new definition of the spin
content of the proton has severe problems with gauge invariance. 

There was one more idea~[8] stating that the EMC result has nothing
to do with the proton spin at all, and should be considered as a
manifestation of some general topological properties of gauge fields.
In the broad sense Ref.~[9] can be also attributed to this trend.

Roughly these three ``schools'' remain to be the main directions of
theoretical interpretation of the puzzle, discovered by the EMC  and
confirmed many times at CERN, SLAC and DESY (HERMES) up to present
time.

We use the term ``puzzle'' instead of often used ``crisis'' or
``problem'' because the EMC result, formally speaking, does not
contradict to the first principles of the official theory of strong
interactions (i.e. QCD). One should note also that at present
nobody is able to calculate measurable polarized (as well as
unpolarized) structure functions, so, in general one has nothing from
the theory to be verified or falsified.

We deal, instead, with some vague symbiosis of QCD and extra
assumptions which can (or cannot) be considered  as ``natural'',
``plausible'' etc.

Nonetheless such a situation seems to be inevitable because one will
hardly be an eye-witness of a beautiful day when it will be possible
to deduce rigorously all or many enough experimentally testable
predictions from the theory (QCD).

The main ideas and tools of the three above-mentioned approaches were
described and discussed in literature and oral talks so many times
that we limit ourselves to their general outline.

The first school based its interpretation of the spin puzzle on the
chiral anomaly, inherent to the singlet axial current and called for
the resque of ``common sense''. ``There is no longer a spin
crisis''~[10]
was a proud conclusion made as early as in~1988 by some proponents of
this approach.

In fact the problem was shifted from the quark part of the proton
spin to its gluon (both spin and orbital) part, the latter being as
ill understood as the quark one. Provided with the ``natural'' value
of the quark contribution (i.e. in full accordance with the quark
constituent model where the nucleon spin is nearly defined by the
quark spin) the adepts of the school under discussion raised the
problem of ``gluonic'' spin contribution both in the sense of its
proper formal definition and calculation and independent experimental
measurement.

The seconds school, as was already said, accepted readily the EMC
data as an exciting evidence that deeply inside the nucleon things
look different from what one could envisage from an earlier
large-distance experience.

There were quite interesting disputes and new proposals (see
e.g.~[11]), which independently of the very subject of discussion
have played very useful r\^ole attracting attention of
theorists to basic notions of theory and motivating new experiments.

At the moment no unique solution of the puzzle in question is
accepted 
by the  ``public opinion'' (I should not like to estimate
percentage).

But I would like to treat the problem from a different point of view.
The matter is that all speculations on the internal content of the
nucleon (i.e. global characteristics such as energy-momentum, spin
etc.) deal with the formal expressions physical interpretation of
which depends on some ``suitable'' convention.

For instance, parton densities in the nucleon are defined by some
composite operators averaged over the 1-nucleon state. These
operators
depend on some arbitrary mass scale (renormalization scale), on one
another arbitrary convention (renormalization scheme). The
interpretation of the matrix elements of these operators as parton
densities depends crucially on the renormalization scheme and, what
is
more serious, on the choice of  a gauge of the gluon field. Such
dependence can be assimilated to frame dependence in special
relativity. This means  that parton interpretation of the matrix
elements of composite operators has no direct objective meaning. In
principle one is able to extract these matrix elements from the data
and check directly the QCD predictions without partonic
interpretation which is akin to mechanical models of electromagnetic
phenomena used in the XIX century.

Somewhat similar approach to the EMC spin effect was undertaken in
Ref.[12], where the value of the integral of $g_1$ was related to
general properties of the gluon field configuration expressed in
terms of the so-called ``topological susceptibility''. General
character of the latter quantity led the authors of this (somewhat
formal) school to
the conclusion about target independence of the EMC spin effect. This
prediction is waiting for the experimental check. But once again its
possible failure will not threat the basics of QCD because the
inference is based on  some extra assumptions wich may or may not
hold in QCD.

Returning to the problem of the parton content of the nucleon spin it
worth noting that disputes on what definition of the ``quark part''
of the nucleon spin is correct cannot be resolved with new
experimental data. Partly because the full $x$-interval $[0,1]$ will
never be covered leaving the room for doubts in non-explored regions
of $x$ (see e.g.~[13]).

But mainly because the ``problem'' being dependent on various
definitions and arbitrary conventions lies in another, metaphysical
plane.

This conclusion can seem quite discouraging and too pessimistic. But
on the other hand this should promote our thoughts about such basic
notions as ``constituents'', conventionality, etc. and their
relevance 
to the interpretation of the data in these terms. In this sense ten
years have passed not for nothing.

At any rate, I believe, it will be interesting for you to know
laconic
characterisations of the present situation made by the two
``culprits''. ``Pregnant!'' (J.~Ellis)~[14];
``Delightful!''(R.~Jaffe)~[15]. I am not pretty sure what they meant,
each of them, but taken together these words are not contradictory
and may promise a birth of new, exciting puzzles.

For, I think, Louis de Broglie was prophetically  right when he wrote
as early as in 1955~[16]: ``... le spin est certainement un des
\'el\'ements les plus essentiels, peut-\^{e}tre m\^eme le plus
essentiel, de l'existence des particules''.

\section*{Acknowledgments}
I am grateful to the Organizing Committee of the International
Symposium ``Spin-98'' for inviting me to give this talk.

\end{document}